\documentclass[11pt]{article}
\usepackage{amsfonts}
\usepackage{amssymb}
\usepackage{amsmath}
\usepackage{amsthm}
\usepackage{latexsym}

\newtheorem{lemma}{Lemma}
\newtheorem{corollary}{Corollary}
\newtheorem{proposition}{Proposition}

\theoremstyle{definition}
\newtheorem{definition}{Definition}

\usepackage{algorithm}
\usepackage[noend]{algpseudocode}

\usepackage{hyperref}
\usepackage{nameref}
\usepackage[left=1.25in, right=1.25in, top=1in, bottom=1in]{geometry}

\title{A note on Graph Automorphism and Smart Reductions}

\author{%
	Eric Allender%
		\footnote{
			Rutgers University, Piscataway, NJ, USA,
			\tt{allender@cs.rutgers.edu}
		} \and \;
  Joshua A. Grochow%
		\footnote{
			University of Colorado at Boulder, Boulder,
                        CO, USA, \tt{joshua.grochow@colorado.edu}
		} \and \;
	Dieter van Melkebeek%
		\footnote{
			University of Wisconsin--Madison, Madison, WI, USA,
			\tt{dieter@cs.wisc.edu}
		}  \and \;
  Cristopher Moore%
		\footnote{
			Santa Fe Institute, Santa Fe, NM, USA, \tt{moore@santafe.edu}
		} \and \;
  Andrew Morgan%
		\footnote{
			University of Wisconsin--Madison, Madison, WI, USA,
			\tt{amorgan@cs.wisc.edu}
    }
}

\newcommand{\promiseZPP}{\mbox{{\sf Promise-ZPP}}}
\newcommand{\promiseBPP}{\mbox{{\sf Promise-BPP}}}
\newcommand{\promiseRP}{\mbox{{\sf Promise-RP}}}
\newcommand{\promiseUP}{\mbox{{\sf Promise-UP}}}
\newcommand{\promisecoRP}{\mbox{{\sf Promise-coRP}}}
\newcommand{\promisecoUP}{\mbox{{\sf Promise-coUP}}}

\newcommand{\ZPP}{\mbox{{\sf ZPP}}}

\newcommand{\RigidGI}{\mbox{\sf Rigid GI}}
\newcommand{\GI}{\mbox{\sf GI}}
\newcommand{\GA}{\mbox{\sf GA}}

\newcommand{\UP}{\mbox{{\sf UP}}}

\newcommand{\coUP}{\mbox{{\sf coUP}}}
\newcommand{\RP}{\mbox{{\sf RP}}}
\newcommand{\coRP}{\mbox{{\sf coRP}}}

\newcommand{\BPP}{\mbox{{\sf BPP}}}

\begin{document}

\maketitle

\begin{abstract}
  It is well-known \cite{kst} that the complexity of the Graph Automorphism
  problem is characterized by a special case of Graph Isomorphism, where the
  input graphs satisfy the ``promise'' of being rigid (that is, having no
  nontrivial automorphisms).  In this brief note, we observe that the
  reduction of Graph Automorphism to the Rigid Graph Ismorphism problem
  can be accomplished even using Grollman and Selman's notion of a
  ``smart reduction''.
\end{abstract}

\section{Prologue}
This paper consists of an orphan theorem.

The history of this work begins with a study of the complexity of
time-bounded Kolmogorov
complexity as it relates to the Graph Automorphism problem \cite{agm}.  The
authors of \cite{agm} continued to explore this topic with other collaborators
after \cite{agm} was posted on ECCC (and on arXiv), and at one point it was found that the
exposition could be simplified by proving that Graph Automorphism can be
reduced to the Rigid Graph Isomorphism promise problem via a smart reduction,
which is the topic of the current note.  The project eventually developed into
a significantly stronger paper \cite{itcs}.  (A more complete version of this
work is available at \cite{agmmm}.)  But the proofs as presented in
\cite{itcs,agmmm} no longer make any reference to smart reductions.

What do do?

The observation that Graph Automorphism reduces to Rigid Graph Isomorphism
might be useful in some future situation, but this fact by itself falls
somewhat short of the Least Publishible Unit threshold.  However, it also
is not compatable with inclusion in \cite{agmmm}.
Since the proof
was already written, and since the availability of this proof might save
someone some effort in the future, it was decided to provide space in this
ECCC/arXiv revision (which also serves the purpose of pointing out that
\cite{agm} has been superceded by \cite{agmmm}) to archive this fact about
smart reductions and Graph Automorphism.

\section{Introduction}

We assume that the reader is already familiar with the Graph Isomorphism problem
($\GI$) and the Graph Automorphism problem $\GA$;
see \cite{kst} for more background.  
It is well-known that $\GA \leq_m^p \GI$ (i.e., $\GA$ Karp-reduces to $\GI$)
but the converse is not known.

A \emph{promise problem} consists of a pair of disjoint subsets $Y,N \subseteq \Sigma^*$ where, as usual, $\Sigma$ is a finite alphabet.  A language $B$ is a
{\em solution} to the promise problem $(Y,N)$ if
$Y \subset B \subset \overline{N}$.  (Note that a language $L$ is simply the
promise problem $(L,\overline{L})$.)
An algorithm $A$
{\em solves} a promise problem $(Y,N)$ if $A(x) = 1$ for all $x \in Y$, and
$A(x) = 0$ for all $x \in N$.
Of particular interest
to us is the promise problem known as the {\em Rigid Graph Isomorphism Problem}.
A graph is \emph{rigid} if it has no nontrivial automorphisms, i.e., none
other than the identity. 
Rigid Graph Isomorphism ($\RigidGI$) is a promise version of $\GI$: namely,
to decide whether two graphs are isomorphic, given the promise that they are
rigid.  That is, $Y$ is the set of pairs of rigid graphs $(G,H)$ such that
$G$ and $H$ are isomorphic, and $N$ is the set of pairs of rigid graphs such
that $G$ and $H$ are not isomorphic.
Thus an algorithm that solves $\RigidGI$ can have arbitrary output if one of
its inputs is not rigid.

We will refer to following ``promise complexity classes''.
\begin{itemize}
\item $\promiseBPP$ is the class of all promise problems $(Y,N)$ for which there
is a probabilistic polynomial-time Turing machine $M$ such that, for all
$x \in Y$, $M$ accepts $x$ with probability at least 2/3, and for all
$x \in N$, $M$ rejects $x$ with probability at least 2/3.
\item $\promiseRP$ is the class of all promise problems $(Y,N)$ for which there
is a probabilistic polynomial-time Turing machine $M$ such that, for all
$x \in Y$, $M$ accepts $x$ with probability at least 2/3, and for all
$x \in N$, $M$ rejects $x$ with probability 1.
\item $\promisecoRP$ is the class of all promise problems $(Y,N)$ for which 
$(N,Y)$ is in $\promiseRP$.
\item $\promiseZPP$ is $\promiseRP \cap \promisecoRP$.
\item $\promiseUP$ is the class of all promise problems $(Y,N)$ for which there
is a nondeterministic polynomial-time Turing machine $M$ such that, for all
$x \in Y$, $M$ has exactly one accepting computation path on $x$, and for all
$x \in N$, $M$ has no accepting computation paths.
\item $\promisecoUP$ is the class of all promise problems $(Y,N)$ for which 
$(N,Y)$ is in $\promiseUP$.
\end{itemize}

An important part of these definitions is that, on inputs outside of $Y\cup N$,
the Turing machine $M$ might not satisfy the ``promise'' (by having acceptance
probability not bounded away from 1/2, or by having more than one
accepting computation path).

The main topic of this note concerns ``smart'' reductions.  In order to
motivate this notion, let us first define what it mans to reduce one
promise problem to another.  Let $(Y',N')$ and $(Y,N)$ be promise
problems.  We say that $(Y',N') \leq_T^p (Y,N)$ if there is a polynomial-time
oracle machine $M$, such that for {\em every}
solution $B$ of $(Y,N)$, the language accepted by $M^B(x)$ is a solution to
$(Y',N')$.  Note that, on any input $x \in Y' \cup N'$, the output of $M$
with oracle $B$ is the same as with any other oracle $B'$ that agrees with
$B$ on $Y \cup N$; any query that is asked by $M$ that lies
outside of $Y \cup N$ can be answered arbitrarily, without affecting the final
outcome, and thus in some sense it is not very ``smart'' for $M$
to bother asking such ``useless'' queries.  This motivated Grolman
and Selman to formulate the following definition:

\begin{definition} \cite{grollman.selman}
A polynomial-time Turing reduction to a promise problem
$(Y,N)$ is called a {\em smart reduction} if it makes queries only to 
elements of $Y \cup N$.
\end{definition}
As Grollman and Selman observe in 
\cite{grollman.selman}, this seems to be a significant
restriction on the space of all possible reductions to promise problems.
In general, reductions to promise problems do not seem to be able to
avoid making ``useless'' queries where the promise does not hold, although
such queries cannot affect the ultimate decision of whether to accept or 
reject.  That is, reductions to promise problems can probably not always
be smart.  (For more on this topic, including applications to the Graph
Isomorphism problem, see \cite{cgrs,glasser.selman.sengupta}.)

As a warm-up, we sketch the known smart reduction of search
to decision for the promise problem of graph isomorphism on rigid
graphs.  (That is, there is a deterministic polynomial-time oracle machine
that, when given two rigid graphs $G_0$ and $G_1$, will either determine
that the graphs are not isomorphic, or else produce an isomorphism between the
two graphs, making only oracle queries to the graph isomorphism problem
where all of the queries consist of pairs of rigid graphs.  This is
completely trivial if the graphs are not isomorphic; thus 
assume that the rigid graphs $G_0$ and $G_1$ are isomorphic. 
There is a (unique) vertex $i$ such that vertex 1 of $G_0$ maps to
vertex $i$ of $G_1$ via an isomorphism.  We can find $i$ using the
decision oracle as follows: For each $j \in [n]$, attach a rigid
``label'' $r$ to vertex $1$ of $G_0$ (call this graph $H$), and attach
the same label to vertex $j$ in $G_1$ (call this graph $H_j$). Note
that both $H$ and $H_j$ are rigid. Query the decision oracle for each
pair $(H,H_j)$, let $i$ be the (unique) $j$ for which the answer
is positive, and set $\pi(1)=i$. We keep the label $r$, i.e., we
continue with $(H,H_i)$, and repeat the process to find $\pi(2)$,
etc. 

We also need the following proposition, which easily follows from
essentially the
same proof as that of \cite{ko2}, showing that if SAT is in $\BPP$, then
it is in $\RP$.)
\begin{proposition}\label{prop:smart}
Let $(Y,N)$ be a promise problem for which there is a smart
reduction from search to decision. Then
$(Y,N) \in \promiseBPP$ implies
$(Y,N) \in \promiseRP$.
\end{proposition}

It is shown in \cite{kst} that $\GA \leq_T^p \RigidGI$, but the reduction
given there is not a smart reduction.  In Section~\ref{mainsec} we
present a smart reduction.  Here, we mention some corollaries that follow
from the existence of a smart reduction, that are not obvious otherwise.

\begin{corollary}\label{easycor}
\begin{itemize}
\item  If $\RigidGI$ is in $\promiseBPP$, then $\GA \in \RP$.
\item  If $\RigidGI$ is in $\promisecoRP$, then $\GA \in \ZPP$.
\item  If $\RigidGI$ is in $\promisecoUP$, then $\GA \in \UP \cap \coUP$.
\end{itemize}
\end{corollary}

\begin{proof}
  The proof of each is similar.  We present only the proof of the final
  implication.  Let $N$ be a nondeterministic machine witnessing that
  $\RigidGI$ is in $\promisecoUP$, and let $M_1$ compute the smart
  reduction from $\GA$ to $\RigidGI$.  On input $G_0$, $M_1$ computes a
  query $(G_1,H_1)$ consisting of two rigid graphs, and asks the oracle if
  $G_1$ is isomorphic to $H_1$.  Our algorithm will guess the answer, and
  verify it by either guessing the unique isomorphism between $G_1$ and $H_1$,
  or else by running $N(G_1,H_1)$, which will have a unique computation path
  if the graphs are not isomorphic.  On the unique branch that verifies that
  the guess was correct, our algorithm will then continue with the simulation
  of $M_1$, to compute the next query $(G_2, H_2)$, and so on.  There will be
  a unique computation path that is able to continue the simulation of
  $M_1(G_0)$ to the end.  To show that $\GA \in \UP$, this unique path will
  accept if and only if $M_1(G_0)$ accepts.  To show that $\GA \in \coUP$, this
  unique path will accept if and only if $M_1(G_0)$ rejects.
\end{proof}

We observe that we do {\em not} know how to prove the implication
``if $\RigidGI$ is in $\promiseUP$, then $\GA \in \UP$.''  This is
especially disappointing, since the hypothesis ``$\RigidGI$ is in $\promiseUP$''
is easily seen to be true.

\section{Main Theorem}\label{mainsec}

\begin{lemma} \label{lem:GA}
There is a smart reduction reducing Graph Automorphism to the 
Rigid Graph Isomorphism Problem.
\end{lemma}

\begin{proof}[Proof of Lemma~\ref{lem:GA}]

  Our proof is patterned after the proof of \cite[Theorem 1.31]{kst},
  which presents a reduction of Graph Automorphism to $\RigidGI$.

Let $G$ be an $n$-vertex graph that is input to the Graph Automorphism
problem.  $G$ has a non-trivial automorphism if and only if there is
an automorphism that sends some vertex $i$ to a vertex $j \neq i$.
Any automorphism fixes some (possibly empty) set of vertices. 

Using the notation of \cite{kst}, let $G_{[j]}^{(i-1)}$ be the graph
(easy to construct in polynomial time, as presented in 
\cite[pages 8 and 31]{kst}) with distinct labels on vertices 
$\{1, \ldots, i-1\}$ (so that no automorphism can move any of those
vertices), and a distinguishing label on vertex $j\geq i$.  
As a slight modification
of this notation, let $G_{[j,k]}^{(i-1)}$ again have distinct labels on 
vertices $\{1, \ldots, i-1\}$ and with two new colors (i.e., labels) $r$ and 
$b$ (red and blue), with $j$ colored $r$ and $k$ colored $b$, where
$j \geq i$ and $k \geq i$.

Let $i$ be the largest index for which some automorphism exists that
fixes vertices $\{1, 2, \ldots, i-1\}$, and sends $i$ to some $j > i$.
Then for some $k>i$, $G_{[i,k]}^{(i-1)}$ and $G_{[j,i]}^{(i-1)}$ are 
isomorphic, and for all $j > i$ and $k > i$, 
$G_{[i,k]}^{(i-1)}$ and $G_{[j,i]}^{(i-1)}$ are rigid (since the first
$i$ vertices have distinct labels).  Furthermore,
for all $\ell>i, G_{[\ell,k]}^{(\ell-1)}$ and
$G_{[j,\ell]}^{(\ell-1)}$ are rigid and non-isomorphic 
for every $j > \ell$ and $k > \ell$.

Thus if we start with $i=n-1$ and pose queries of the form
$(G_{[i,j]}^{(i-1)},G_{[k,i]}^{(i-1)})$ (for $j,k \in \{i+1, \ldots, n\}$)
to an oracle for the Rigid Graph Isomorphism Problem, it holds that
for all large values of $i$ the graphs are rigid and non-isomorphic (and
thus satisfy the promise of the promise problem $(Y,N)$, until we encounter 
the first triple $(i,j,k)$ such that the graphs
$(G_{[i,j]}^{(i-1)},G_{[k,i]}^{(i-1)})$ are isomorphic. These graphs
are also rigid, and thus they also 
satisfy the promise.
If the computation ends with all queries determined to be non-isomorphic,
then this is a proof that $G$ has no nontrivial automorphism.

This algorithm works correctly on all inputs, 
and all queries satisfy the promise.  Thus it is a smart reduction.
\end{proof}

Since the ``promise'' in the $\RigidGI$ promise problem is precisely the
problem solved by $\GA$, it is perhaps worthwhile to observe that
hypotheses regarding the complexity of the promise problem $\RigidGI$
yield conclusions about rigid graph isomorphism that do not need to be
phrased in terms of promise problems:

\begin{corollary}
Let 
$A$ = \{$(G,H)$ : $G$ and $H$ are rigid, and $G$ is isomorphic to $H$\}, and

$B$ = \{$(G,H)$ : $G$ and $H$ are rigid, and $G$ is not isomorphic to $H$\}.
\begin{itemize}
\item  If $\RigidGI$ is in $\promisecoRP$, then $A$ and $B$ are in $\ZPP$.
\item  If $\RigidGI$ is in $\promisecoUP$, then $A$ and $B$ are in
  $\UP \cap \coUP$.
\end{itemize}
\end{corollary}

\begin{proof}  The proof of each part is similar.  Assume that $\RigidGI$
  is in $\promisecoRP$.

The first step is to determine if an input pair
$(G,H)$ consists of two graphs, {\em both} of which are rigid.
But testing if one of $\{G,H\}$ has a
non-trivial automorphism can be determined in $\ZPP$,
by Corollary~\ref{easycor}.

Thus the algorithm is as follows:
On input $(G,H)$ determine (in $\ZPP$)
whether $G$ and $H$ are both rigid.

If both are rigid, run the $\promisecoRP$ algorithm to attempt to find a
proof that $G$ and $H$ are not isomorphic.  (This will show that
$\overline{A}$ and $B$ are in $\coRP$.)  Or, one can run the
$\promiseRP$ algorithm for $\RigidGI$ (as guaranteed by
Proposition~\ref{prop:smart}) to attempt to find a proof that $G$ and
$H$ are isomorphic.  (This will show that
$A$ and $\overline{B}$ are in $\coRP$.)
The promise is satisfied, so the probability of success is 
guaranteed to be high.

For the other implication, the hypothesis implies $\GA \in \UP \cap \coUP$.
Thus, on input $(G,H)$, there is a unique computation path that either
provides a proof that one of $G$ and $H$ is not rigid, or else provides a
proof that both are rigid.  To show that $A$ and $\overline{B}$ are in
$\UP$, one can guess the unique isomorphism between $G$ and $H$.
To show that $\overline{A}$ and $B$ are in $\UP$, one can use the
hypothesized $\promisecoUP$ algorithm for $\RigidGI$.
\end{proof}

\subparagraph*{Acknowledgments.}
E.~A. acknowledges the support of National Science Foundation
grant CCF-1555409.
J.~A.~G. was supported by an Omidyar Fellowship from the Santa Fe Institute and mNational Science Foundation grant DMS-1620484.
D.~v.~M. acknowledges the support of National Science Foundation
grant CCF-1319822.
We thank V.~Arvind for helpful comments about the graph automorphism
problem and rigid graphs.

\bibliographystyle{alpha}
\bibliography{refer}

\end{document}